\documentclass[%
 reprint,
 amsmath,amssymb,
 aps,
]{revtex4-2}

\usepackage{graphicx}
\usepackage{dcolumn}
\usepackage{bm}
\usepackage{xcolor,soul,framed} 
\usepackage{subfigure}
\usepackage{float}
\usepackage{afterpage}

\begin{document}


\title{Gain and One-Way Propagation in \\ Synthetically Moving non-Foster Gratings}


\
\author{Antonio Alex-Amor}
\affiliation{
Department of Electronic and Communication Technology, RFCAS Research Group, \\  Universidad Autónoma de Madrid, 28049 Madrid, Spain
}%
\author{Carlos Molero}
\affiliation{%
Department of Electronics and
Electromagnetism, Physics Faculty, Universidad de Sevilla, \\
Avenida de la Reina Mercedes S/N, 41012 Sevilla, Spain.
}%
\author{Mário G. Silveirinha}
\affiliation{
Instituto Superior Técnico and Instituto de Telecomunica\c{c}oes, University of Lisbon, 1049-001 Lisboa, Portugal. 
}%

\begin{abstract}
In this paper, we analyze the electromagnetic properties of space-time grooved metal surfaces governed by uniform-velocity modulations. We begin by deriving the electromagnetic fields of a uniform-velocity-modulated parallel-plate waveguide (PPW) using Lorentz transformations, establishing it as the fundamental building block of a more complex space-time structures. We then analyze the dispersion and scattering characteristics of a space-time grooved surface and later extend the study to the interaction between two facing surfaces. Our findings show that these systems can indeed amplify electromagnetic waves and exhibit non-reciprocal as well as non-Foster behavior. Moreover, under specific conditions, they enable the formation of unidirectional propagation channels, effectively constraining light to be guided along a single direction. These results unveil new opportunities for the design of advanced electromagnetic and photonic devices.
\end{abstract}

\maketitle



\section{\label{sec:introduction}Introduction}

Time has recently been explored as a new degree of freedom for controlling wave propagation in electromagnetic systems \cite{engheta2023four,  Sounas2017, TaravatiCaloz2017, ZhanCui2018,  taravati2022microwave, Tirole2023double, TretyakovAsadchy2023, Fang2023multifunctional, Lustig2023, Pacheco2025, SilveirinhaPlasmonicTime2025}. Such systems are attractive because time modulation of the material response can serve two purposes: (i) it can break time-reversal invariance, which is linked to reciprocity; and (ii) it can break time-translation symmetry, which is associated with energy conservation. In particular, time-varying systems have been proposed as a potential route to introducing gain into the optical response \cite{Pendry2021, Lyubarov2022PTC, LiberalAmplification2023}. Under suitable conditions, the modulation can continuously transfer energy into the electromagnetic field, leading to regimes of parametric instability in which the field amplitude grows progressively over time.

Such instability regimes have recently attracted considerable attention in space-time-modulated systems \cite{Zurita2009, Fleury2018, Galiffi2022_PTVM, Serra2024ParticleHole}, where spatial and temporal modulations combine to form crystalline-like electromagnetic structures. A paradigmatic example is a space-time crystal with a permittivity that varies sinusoidally in both space and time \cite{Oliner1961, Cassedy1963}, thereby creating a traveling-wave modulation propagating through the medium. The modulation front behaves analogously to a moving medium and can induce Fresnel-drag-like interactions with electromagnetic waves, closely resembling the physics of physically moving bodies \cite{Huidobro2019, PrudencioMinkowskian}. Depending on the relation between the effective modulation velocity and the phase velocity of light in the medium, the system may operate in superluminal or subluminal regimes \cite{spacetime_caloz1, spacetime_caloz2}. In these regimes, Doppler-shifted mode coupling enables non-reciprocal energy transfer between the waves and the traveling modulation, providing the physical mechanism underlying amplification and instability.

Predictions of instabilities in systems under space-time modulation date back to the 1960s \cite{Cassedy1967}. Essentially, the space-time modulation emulates a traveling-wave front \cite{Cullen1958}, whose synthetic velocity $v$ is defined by the ratio of the temporal to the spatial modulation frequencies \cite{Galiffi2019}.
The modulation is considered superluminal when the effective velocity exceeds the characteristic (phase) velocity of the medium $v_{\text{c}}$ ($v > v_{\text{c}}$), with $v = v_{\text{c}}$ marking the Cherenkov (or luminal) threshold \cite{Oue2021PHD}.
This class of systems is part of a broader framework involving synthetic moving modulations \cite{Harwood2025,taravati2019}, where the resulting space-time gratings may exhibit gain, as well as other remarkable phenomena such as analogies to Hawking radiation \cite{Horsley2023PNAS} and black-hole physics \cite{Dentcho2009}. 

This article considers  uniform-velocity modulations, which produce  close analogues of moving media in which one of the components physically moves at a constant velocity \cite{Yeh1966}. In fact, moving systems can themselves be regarded as particular cases of space-time modulated platforms. Physically moving gratings exhibit Doppler shifts \cite{Dossou2016}, and may induce non-reciprocal and frequency-mixed diffraction patterns \cite{Kelly1982}, among other phenomena. 

Furthermore, closely spaced moving bodies undergoing shear motion have also been linked to electromagnetic instabilities \cite{Maslovski2013, Silveirinha2014, Oue2025, Brevik2022}. In particular, the intrinsic connection between the emergence of instabilities and the Cherenkov threshold, a velocity limit that marks the boundary between stable and unstable systems, has been highlighted in different works \cite{Maghrebi2013, Maslovski2013, Silveirinha2014, Lannebere2016}. These instabilities typically manifest as signal amplification and are often associated with the hybridization of guided modes supported by moving objects, which can give rise to unidirectional propagation channels. Furthermore, such instabilities, specifically the gain response at their origin, have also been linked to the phenomenon of quantum friction \cite{Pendry1997, Pendry2010, Silveirinha2014QF, Oue2024, VolokitinRMP}. The implications of this force, which has provoked passionate and thought-provoking debates \cite{Philbin2009, Pendry2010, Leonhardt2010, Pendry2010_2}, have generally been explored within the paradigm of quantum electrodynamics. 

Semiconductors biased with drifting electrons constitute a second class of  analogous systems involving effectively moving entities  \cite{Sumi1967, Riyopoulos2009}.  
They exhibit related electromagnetic (or optical) instabilities in the form of amplification, including scenarios operating in the THz regime \cite{Sydoruk2010}. The net gain reported in these works is based on the assumptions outlined in \cite{Solymar1966}, where the instability primarily arises from the interaction between drifting electrons and the guided modes supported by the system. Guiding structures incorporating graphene layers have also been studied under DC-current pumping, due to the high electron mobility in graphene \cite{Shishir2009, Morgado2018, Morgado2020, Morgado2022}. Instabilities may emerge when drifting electrons interact with surface plasmons \cite{Aryal2016, Dadoenkova2017, Dadoenkova2018}. The Cherenkov threshold is essentially surpassed for plasmons with sufficiently short wavelengths, leading to an effective negative conductivity that results in net optical gain and one-way propagation \cite{Morgado2017,Morgado2021}.

In this work, inspired by the rich phenomenology of interacting moving bodies, we explore space-time variant electromagnetic systems that can be pictured as uniform-velocity grooved-metal surfaces. As the slits of the structure are sufficiently thick, the unit-cell analysis emulates a parallel-plate waveguide (PPW) problem \cite{Rui2011, Medina2010, Molero2016, Molero2014, Rotman1963}. 

The electromagnetic analysis of uniform-velocity space-time crystals can be conveniently addressed in comoving coordinate frames, involving static/stationary materials \cite{PrudencioMinkowskian}. Although moving dielectrics are more subtle to treat because they become bianisotropic,  \cite{paloma_spacetime, PrudencioMinkowskian, Mirmoosa2024, Prudencio2024}, special gratings formed only by infinitesimally thin air–metallic insertions \cite{Berral2012,  Alex2023TV, SalvadorAccess2023, MorenoRodriguez2024}, maintain their electrical properties regardless the coordinate frame. 
In particular, we recently studied a moving slit grating under plane wave illumination \cite{AlexAmor2023}, and demonstrated how the synthetic motion induces asymmetric Goos-Hanchen shifts, asymmetric scattering patterns, among others.

Here, we apply the same strategy to characterize the electromagnetic response of the space-time grooved surface. Importantly, unlike in our previous work \cite{AlexAmor2023}, the response of the grating is no longer frame-independent due to the finite thickness of the system along the direction perpendicular to the grating plane. Hence, the problem represents a nontrivial extension of our previous theory.

We develop a homogenization model that describes the system in terms of an equivalent \emph{space-time surface impedance}, and demonstrate that the effective response violates Foster's theorem \cite{Foster1924} and can enable negative-energy regimes. 
The transformation of the  surface impedance in the comoving frame to the laboratory frame is performed using the expressions derived in \cite{AlexAmor2023}. 
Inspired by the physics of moving systems, we characterize the guided modes supported by the considered space-time-modulated surfaces, and identify one-way propagation regimes. Furthermore, we study the interaction of two such systems modulated at different synthetic velocities and reveal phenomena that can be regarded as classical analogues of quantum friction and spontaneous light emission \cite{Silveirinha2014}.

The paper is organized in the following way: Section II presents the theoretical description of the uniform-velocity-modulated PPW, which establishes the foundations and analytical expressions to be used in the rest of the paper. Sections III and IV focus on a single grooved surface, providing the analytical framework for the homogenization that yields the effective surface impedance, reflection coefficient, and guided modes. Section V describes the configuration based on the two facing grooved metal surfaces, in which unstable one-way propagation behavior emerges.     

\section{Space-Time Parallel-Plate Waveguide}

\begin{figure}[!t]
	\centering	
    \subfigure[]{\includegraphics[width=0.6\columnwidth]{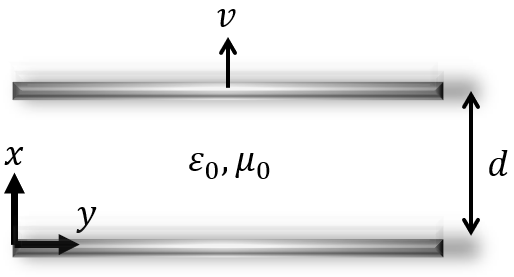}} 
    \subfigure[]{\includegraphics[width=0.9\columnwidth]{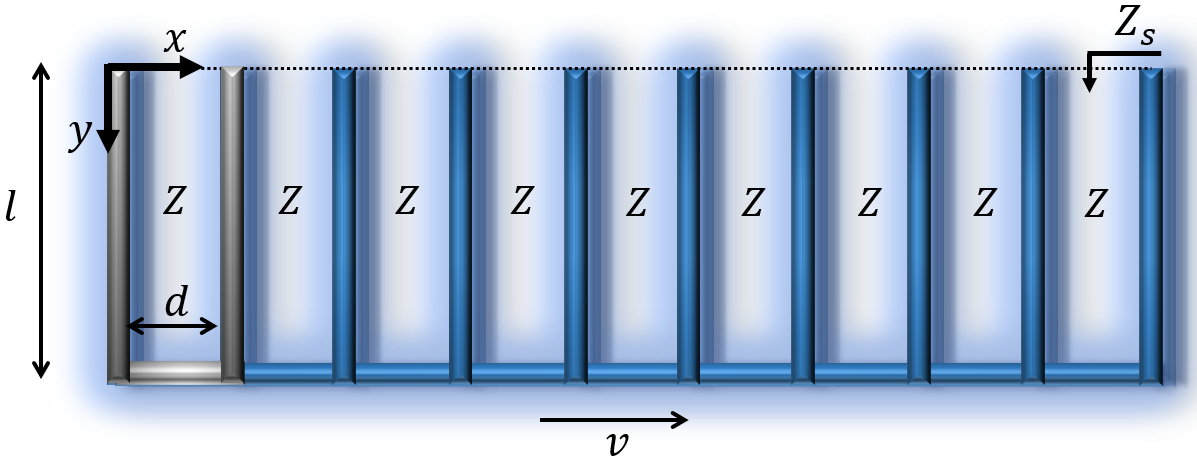}} 
	\caption{Sketch of space-time-modulated metallic metastructures. (a) Parallel plate waveguide (PPW). (b) Grooved space-time metal surface. The system is formed by periodic repetitions of short-circuited uniform-velocity-modulated PPWs. The whole metastructure can be equivalently seen as a space-time surface impedance $Z_s$.} 
	\label{bednails}
\end{figure}

As Figure \ref{bednails}(b) illustrates, the structure under analysis consists of a uniform-velocity (synthetic speed $v$)  grooved metal surface that repeats periodically along the $y$-axis. The grooves depth is $l$ and the metal walls thickness is assumed infinitesimal. The grooved metal surface can also be regarded as periodic array of parallel-plate waveguides (PPW) whose plates are separated a distance $d$. 

In order to describe the physics of the grooved surface, first we will study the response of a single moving PPW [Figure \ref{bednails}(a)]. Thus, first we will consider that the two metallic plates that form the PPW are infinitely extended along the  $y$ and $z$ directions. The PPW  ``moves" with a uniform velocity $v$ in the $x$ direction.  Moreover, the PPW walls are perfect electric conductors (PEC) and the dielectric is air ($\varepsilon_r = \mu_r = 1$).

We assume that in the laboratory frame, the tangential electric field vanishes at the metallic plates of the time-varying PPW, i.e., $\hat{\mathbf{n}} \times \mathbf{E} = \mathbf{0}$ at $x=0,d$. The parameter $d$ is the distance, in the laboratory frame, between the two metallic plates that form the PPW. Importantly, the boundary condition in the comoving frame is different from that in the laboratory frame. In fact, using a Lorentz transformation of the electromagnetic field, one can find that in the comoving frame: 
\begin{equation} \label{BC_comoving}
    \hat{\mathbf{n}} \times(\mathbf{E}'-\mathbf{v} \times \mathbf{B}') = \mathbf{0}\, , \quad \text{at}\,\, x'=0,d',
\end{equation}
where $d' = \gamma d$ is the dilated distance between metallic plates, with $\gamma = 1 / \sqrt{1 - (v/c)^2}$ being the Lorentz factor. Note that the unprimed notation in the electromagnetic fields, space-time coordinates, and spectral parameters refers to the original laboratory frame, while the primed notation ($'$) to the static comoving frame. Lorentz transformations are summarized in the Appendix.

The dependence of the boundary condition \eqref{BC_comoving} on the modulation speed $v$ is a fingerprint of bianisotropy, a characteristic feature of space-time-modulated systems in their comoving frame \cite{PrudencioMinkowskian}. It is important to emphasize that, because the boundary condition in the comoving frame does not reduce to the standard PEC condition, our space-time surface is \emph{not} equivalent to a physically moving grating, in contrast to the case discussed in our previous work \cite{AlexAmor2023}. Indeed, the PEC boundary condition is satisfied in the laboratory frame, where the system is time-varying, rather than in the frame where the medium response is time-independent. Moreover, we shall see below that the boundary condition \eqref{BC_comoving} is non-Hermitian, implying that the boundary wall can either generate gain or absorb energy.

Now, we particularize the analysis to TM incidence. The modal solutions excited by this type of incidence include the TEM mode, which propagates without a cut-off frequency and is responsible for much of the observed phenomenology. TM waves are described by fields with components $E_x, E_y$ and $H_z$. The vector boundary condition \eqref{BC_comoving} in the comoving frame reduces to the scalar equation
\begin{equation} \label{BC_comovingTM}
   E'_y + v\mu_0 H'_z=0\, , \quad \text{at}\,\, x'=0,d' \,.
\end{equation}
which is a consequence of $E_y = 0$ at $x = 0, d$ in the laboratory frame. We look for modal solutions with the structure $\mathbf{H}' = H_z'(x',y')\, e^{j\omega' t'} \hat{\mathbf{z}}$, of time-harmonic nature. As the fields in the Lorentz-comoving frame satisfy the free-space Maxwell's equations in the air region (note that air, modeled here as vacuum, is invariant under a Lorentz transformation), $H_z'(x',y')$ has to satisfy the Helmholtz equation,
\begin{equation}
\left(\frac{\partial^2}{\partial x'^2} +
\frac{\partial^2}{\partial y'^2}\right) H_z' + \left(\frac{\omega'}{c}\right)^2 H_{\text{z}}'=0\,,
\end{equation}
whose general solution is
\begin{equation} \label{Hzprime_TM}
    H_z'(x',y')=e^{-j k'_y y'}\left[A \cos \left(k'_x x'\right)+B \sin \left(k'_x x'\right)\right]\,,
\end{equation}
with $A$, $B$ being integration constants and 
\begin{equation} \label{kyprime}
    k_y' = \sqrt{\left(\frac{\omega'}{c} \right)^2 - k_x'^2}\,,
\end{equation}
the propagation constant in the comoving frame.

The electric field in the comoving frame can be obtained from Ampère's equation ($\nabla \times \mathbf{H'} = j\omega' \varepsilon_0 \mathbf{E'}$) and $H_z'$ in \eqref{Hzprime_TM}:
\begin{align} \label{Efieldx}
    E_x' &= -\frac{k_y'}{\omega'\varepsilon_0}\, H_z' \\  
    \label{Efieldy}
    E_y' &= j\frac{1}{{\omega'\varepsilon_0}}\,\frac{\partial H_z'}{\partial x'} 
\end{align}

In particular, from the above formulas, the scalar boundary condition in \eqref{BC_comovingTM} can be rewritten as
\begin{equation} \label{BC_comovingTM2}
    -\frac{\partial H_z'}{\partial x'}  +j \frac{v \omega'}{c^2} H_z'=0\, , \quad \text{at}\,\, x'=0,d'\,.
\end{equation}

Enforcing the boundary condition \eqref{BC_comovingTM2} on \eqref{Hzprime_TM}, we find
\begin{multline} \label{BC_comovingTM3}
    A\left(j \frac{v \omega'}{c^2} \cos \left(k_x' x'\right) + k_x' \sin \left(k_x' x'\right)\right)
    \\+
    B\left(j \frac{v \omega'}{c^2} \sin \left(k_x' x'\right) - k_x' \cos \left(k_x' x'\right)\right)=0\,,
\end{multline}
which has to be satisfied at positions $x' = 0, d'$. Solving Eq. \eqref{BC_comovingTM3} for $x'=0$ and $x'=d$ leads to
\begin{equation}
    B = A  \frac{j v \omega'}{k_x' c^2}\, ,
\end{equation}
\begin{equation} \label{BC_comovingTM4}
    \sin \left(k_x' d'\right)\left[k_x'^2 - \left(\frac{v \omega'}{c^2}\right)^2 \right]=0\,.
\end{equation}

\begin{figure}[!t]
	\centering	\subfigure{\includegraphics[width=1\columnwidth]{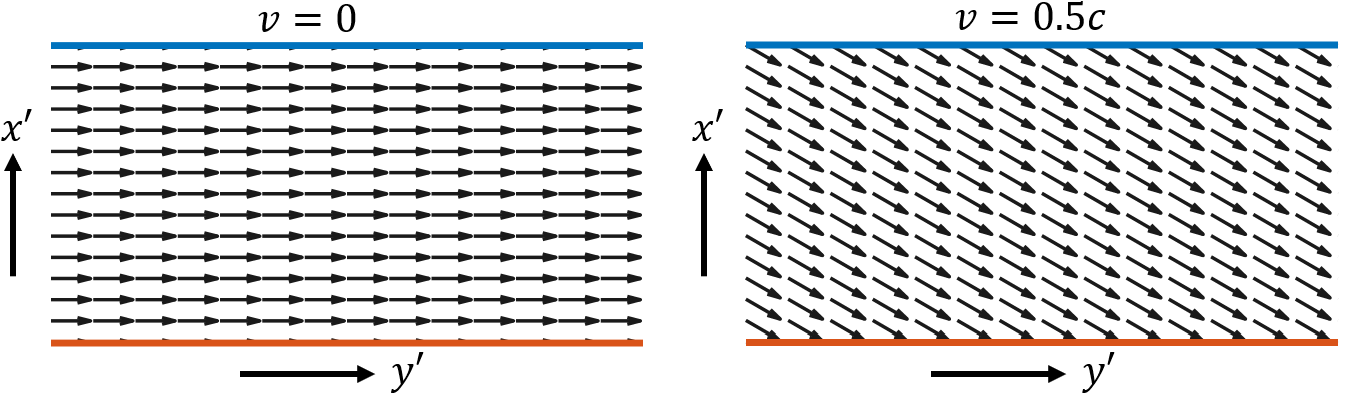}
	} 
	\caption{Poynting vector lines in the comoving frame for the fundamental TM mode of the time-varying PPW.  Two different time modulations are considered: $v=0$ (left panel) and $v=0.5c$ (right panel). Blue and red lines mark the upper and lower metallic plates, respectively. } 
	\label{Poynting}
\end{figure}

From \eqref{BC_comovingTM4}, we obtain solutions of the type
\begin{equation} \label{kx_prime1}
    k_x' d' = n\pi, \quad n=1,2,...
\end{equation}
and, the peculiar solution
\begin{equation} \label{kx_prime2}
    k_x' = \pm \frac{v \omega'}{c^2}.
\end{equation}
We are especially interested in the latter, as it leads to a TEM mode in the laboratory frame, as discussed in the following.  Replacing \eqref{kx_prime2} in \eqref{Hzprime_TM}, we obtain
\begin{equation} \label{Hzprime_co}
    H_z'(x',y') = A\, \mathrm{e}^{j|k_x'| x'}\, \mathrm{e}^{-j k_y' y'}, 
\end{equation}
and, subsequently, the expressions in \eqref{Efieldx}-\eqref{Efieldy} read (the time dependence $\text{e}^{j \omega' t'}$ is suppressed)
\begin{align}
E_{x}'(x', y') &= -A\frac{k_{y}'} {\omega ' \varepsilon_{0}}\,  \mathrm{e}^{j|k_x'| x'}\, \mathrm{e}^{-j k_y' y'}, \\
E_{y}'(x', y') &= -A\frac{|k_{x}'|} {\omega ' \varepsilon_{0}} \,  \mathrm{e}^{j|k_x'| x'}\, \mathrm{e}^{-j k_y' y'},
\end{align}
with $k_x'$ given in \eqref{kx_prime2} and propagation constant $k_y'$:
\begin{equation} \label{kyprime2}
    k_y' = \frac{\omega'}{c} \sqrt{1 - \left(\frac{v}{c} \right)^2} = \frac{\omega'}{\gamma c}\, .
\end{equation}
 This solution represents a TEM-like wave propagating obliquely in the comoving frame, with angle
\begin{equation} \label{theta_prime}
    \theta' = \arctan (v\gamma/c), \,
\end{equation}
measured with respect to the $y'$ axis. Thus, if the PPW is not modulated ($v=0$), $E_y'$ vanishes and the Poynting vector is parallel to the metallic plates. However, if the the PPW is modulated ($v\neq 0$), the Poynting vector seen in the comoving frame emerges from one of the plates and ends on the opposite plate. This is illustrated in Figure \ref{Poynting} for the modulation speeds $v=0$ and $v=0.5c$. Note that, according to Eq. \eqref{theta_prime},  the Poynting vector is tilted by $\theta' = 30$ deg when $v=0.5c$.

As the Poynting vector in the comoving frame has a component along $x'$, it follows that one of the plates generates energy, while the other plate absorbs it. Thus, the Poynting vector lines reveal the non-Hermitian (active) nature of the space-time-modulated system. Interestingly, as the amount of energy emitted by one plate is exactly equal to the amount of energy absorbed at the other plate, the mode energy density remains independent of time.

Switching back to the original laboratory frame, by applying an inverse Lorentz transformation of the comoving fields (see the Appendix) we get that $E_y = 0$, $E_x \sim e^{-j \frac{\omega}{c} y} e^{j \omega t}$ and $H_z =- E_x/\eta_0$, with $\eta_0$ being the free-space impedance.
This result agrees with the expression of a standard TEM mode in the laboratory frame, but which is dragged in space along the $x$ direction due to the synthetic motion. In fact, the field lines of the static TEM wave, with an electric field perpendicular to the plates, are compatible with the PEC boundary condition. In other words, the TEM mode dispersion in the laboratory-frame
coordinates is unaffected by the space-time modulation.

For the TEM-like solution, the characteristic impedance of the time-varying PPW in the comoving frame can be calculated  as
\begin{equation} \label{Z_TM_comoving}
    Z' = \frac{-E_x^{\prime}}{H_z^{\prime}} = \frac{k_y'}{\omega' \varepsilon_0} 
\end{equation}
Noting that $k_y' = \omega' / (\gamma c)$ [Eq. \eqref{kyprime2}], the expression for the characteristic admittance in the comoving frame can be simplified to
\begin{equation} \label{Zprime_TM}
   Z' = \frac{1}{\varepsilon_0 c \gamma} =  \frac{Z_0}{\gamma}
\end{equation}
Hence, the characteristic impedance is frequency independent and gets smaller when the synthetic velocity of the PPW increases. In the limit $v=c$, the characteristic impedance $Z'$ becomes zero. At low modulation speeds ($v\ll c$), $Z'$ can be approximated by the free-space impedance $Z_0$. 

The analyzed TEM-like solution represents a propagative mode with no  cut-off [see Eq.~\eqref{kyprime2}]. In contrast, higher-order modes have a cut-off frequency, which marks their propagation threshold. It follows from Eq. \eqref{kx_prime1}  that the comoving frame cut-off frequency of the $n$th-order mode is
\begin{equation}
\omega_{\text{c},n}' = \frac{n\pi c}{d'} \,.
\end{equation}
The cut-off frequency associated with the first higher-order harmonic, $n = 1$, establishes the upper frequency limit below which the fundamental mode propagates alone. For frequencies $\omega' < \omega'_{\text{c},1}$, higher-order modes can be neglected, and each PPW can be modeled as a transmission line. In the following section, we use this property to derive the effective response of the space-time grooved metal-surface.

\section{Homogenization of the Space-Time Grooved Surface}

Now that the modes of the space-time-modulated PPW have been found, we can characterize the space-time periodically grooved  surface of Figure \ref{bednails}(b). Each groove can be regarded as a truncated PPW with depth $l$. The grooves are terminated by a metallic plate. The metallic walls that delimit the grooves move effectively  along the $x$ axis with uniform speed $v$. As before, the distance between the plates of each elementary PPW is $d$. 

We derive an effective description of the system in term of surface impedance $Z_s$. Our model assumes that the waveguides are densely packed ($d \ll l$), so that a single mode description within each PPW becomes feasible. Thus, we can directly focus on the lower-order (dominant) mode with $k_x'$ given by Eq. \eqref{kx_prime2}.

The effective impedance $Z_s$ is independent of space (coordinate $x$) and time. Thus, it will be useful to characterize the electromagnetic response of the system, such as its gain/dissipative properties, the reflection coefficient, or the dispersion of the guiding modes at the interface air/grooved-surface. 

\subsection{Equivalent Space-Time Surface Impedance}

For densely-packed array of PPWs ($d \ll l$), the grooved surface can, under TM incidence, be approximated in the comoving frame as
\begin{equation} \label{Zs_comoving}
    Z_s' \approx j Z' \tan(k_y' l) = j \frac{Z_0}{\gamma} \tan \left( \frac{\omega' l}{\gamma c} \right)\, .
\end{equation}
This formula is obtained using standard transmission line theory, noting that each PPW waveguide is terminated in short-circuit. Note that the direction of the effective motion is parallel to the termination plate. For this reason, the boundary condition at the termination plate remains that of a PEC in the comoving frame, unlike the plates of each PPW, which, because of their different orientation, are characterized by a frame-dependent boundary condition [Eq. \eqref{BC_comoving}]. Consequently, the equivalent circuit of the termination plate remains a short circuit in the co-moving frame.
In the above, $Z'$ [Eq. \eqref{Z_TM_comoving}] denotes the characteristic impedance of the dominant TEM mode in the comoving frame, and $k'_y$ [Eq. \eqref{kyprime2}] the corresponding propagation constant. 

The surface admittances in the laboratory and comoving frames are related as \linebreak $Y_s = Y_s' / (\gamma [1-k_xv/\omega])$ \cite{AlexAmor2023}. Using this result, we find that the surface impedance in the laboratory frame can be expressed as
\begin{equation}
     Z_s = j Z_0 \left(1- \frac{vk_x}{\omega} \right) \tan \left( \frac{\omega' l}{\gamma c} \right)\,.
\end{equation}
Using now the relativistic Doppler shift formula \linebreak $\omega' = \gamma (\omega - vk_x)$, the surface impedance $Z_s$ can be expressed as a function of only the laboratory-frame parameters:
\begin{equation} \label{Zs_TM}
    Z_s = j Z_0 \left( \frac{\omega - vk_x}{\omega} \right)  \tan \left( \frac{(\omega - vk_x) l}{c} \right)\, .
\end{equation}

The homogenization is expected to be valid provided that i) the array is densely-packed ($d\ll l$), so only the TEM-like mode is relevant and the higher-order modes can be neglected; and ii) $\omega' d' /c \ll 1$ and $k_x' d'/c \ll 1$, which are the usual homogenization constraints. Note that except in the ultra-relativistic  regime, where the $\gamma$ factor deviates significantly from unity, the prime symbols in condition ii) can be dropped.

For future reference, we note that Eq.~\eqref{Zs_TM} can  be further simplified when operating in the quasi-static (low-frequency, long-wavelength) regime. In the quasi-static regime, the tangent function can be approximated by its argument. As a consequence,
\begin{equation}
    Z_s \approx j \frac{Z_0 l}{c} \frac{(\omega - vk_x)^2}{\omega}, \quad \text{quasi-static} .
\end{equation}

\subsection{Non-Foster Response and Negative Stored Energy}

In the general case, the reactance $X$ that models the spacetime grooved metal surface is ($Z_s = j X$) 
\begin{equation} \label{reactance}
    X = Z_0 \left( \frac{\omega - vk_x}{\omega} \right)  \tan \left( \frac{(\omega - vk_x) l}{c} \right)\, .
\end{equation}
According to the Foster's reactance theorem \cite{Foster1924}, a passive and weakly dissipative system is constrained by $\partial X / \partial \omega > 0$ \cite{Serra2023, AlexAmor2025, NonFosterTime_Ptitcyn2023, NonFosterTime_Pacheco2025, NonFoster_Ziolkowski2012}. In words, the reactance of a weakly dissipative system must increase monotonically in frequency.  This condition is satisfied by our model when $v = 0$, because in that case \linebreak $\partial X / \partial \omega = Z_{0}  l[c\cos^{2}(\omega l /c)]^{-1} > 0$. However, for the space-time-modulated structure, this condition may fail in the laboratory frame because such systems are non-conservative, i.e., their physics is governed by processes that involve both dissipation and gain \cite{Serra2023}.

\begin{figure}[!t]
	\centering	\subfigure[]{\includegraphics[width=0.9\columnwidth]{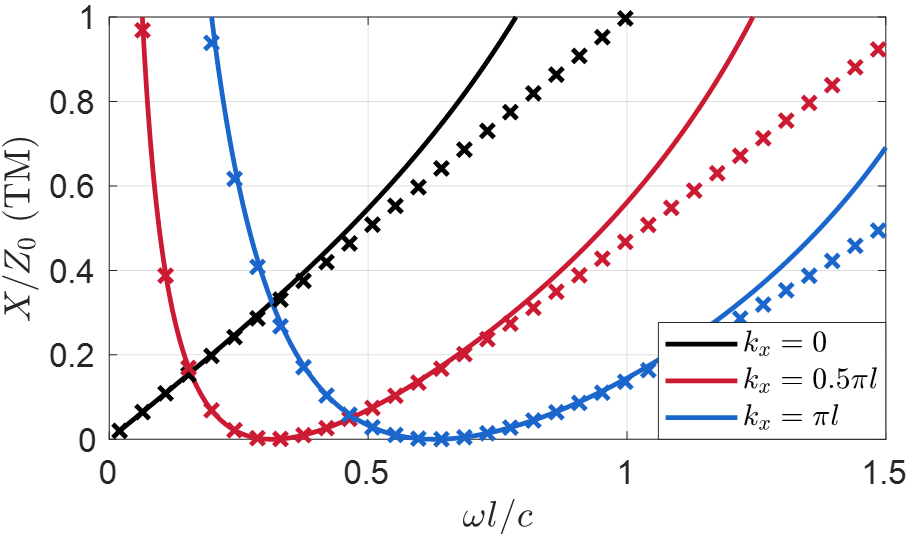}}
    \subfigure[]{\hspace{-0.2cm}\includegraphics[width=1.02\columnwidth]{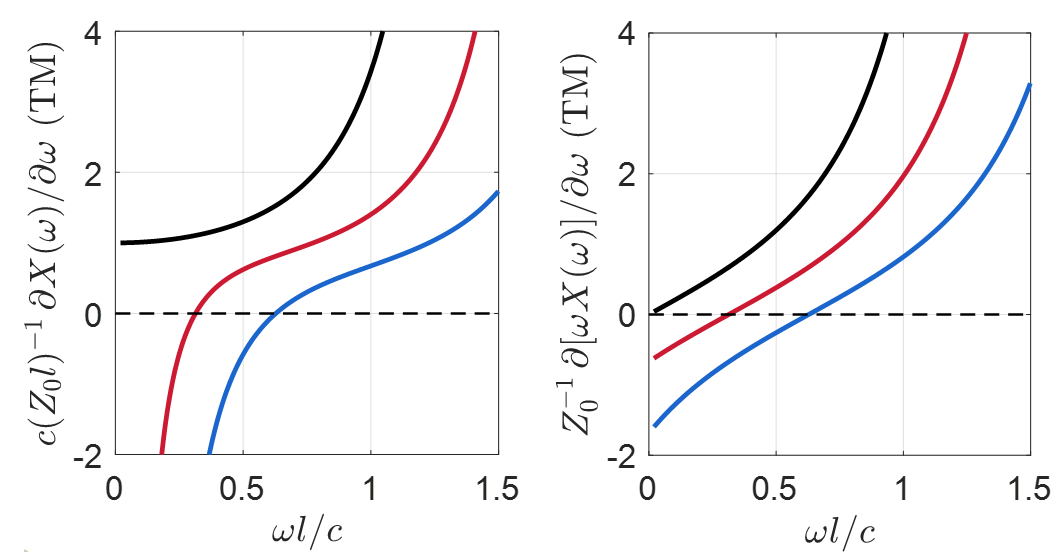}}
	\caption{(a) Normalized reactance $X/Z_0$ as a function of the normalized frequency $\omega l /c$ for different values of $k_x$ and modulation velocity $v=0.2c$ under TM incidence. Solid lines represent the full reactance, while crosses represent its quasi-static approximation valid in the low-frequency regime. \linebreak 
    (b, left panel) Normalized frequency derivative of the reactance, $\partial X / \partial \omega$. (b, right panel) Normalized frequency derivative of the frequency-reactance product, $\partial [\omega X] / \partial \omega$. Regions where $\partial X / \partial \omega <0$ and $\partial [\omega X] / \partial \omega < 0$ are associated with non-Foster behavior and negative stored energy, respectively. } 
	\label{fig:reactance}
\end{figure}

The normalized reactance $X/Z_0$ is plotted as a function of the frequency $\omega$ in Figure \ref{fig:reactance}(a). As seen, the reactance does not exhibit a monotonic variation with frequency for a
fixed $k_x$ and sufficiently small frequencies ($\omega/c \ll |k_x|$), and thus exhibits a \emph{non-Foster behavior}. This can be better appreciated in Figure \ref{fig:reactance}(b, left panel) through the regions where $\partial X / \partial \omega <0$. In fact, the tendency of the curves manifests a negative evolution from $\omega = 0$ to the inflection point ($\omega = 0.3c/l$ for $k_{x} = 0.5\pi/l$), which is the region where the derivative is negative. In general, the spectral region with non-Foster behavior corresponds to the range where $\omega-k_x v< 0$, or equivalently \linebreak $k_x > \omega/v$, which is typically associated with waves that exhibit rapid spatial variations.

Furthermore, as is well known, for passive and weakly dissipative systems, 
$\partial [\omega X(\omega)]/\partial \omega$ governs the energy stored in the grooved surface 
region in the time-harmonic regime. In particular, this quantity must be 
positive to ensure a positive stored energy, understood as the energy transferred 
by an incoming wave to the grooved surface region during the transient process before the steady state is 
reached \cite{Serra2023}. However, non-Hermitian platforms, such as space-time-modulated systems, 
are not subject to the same positive-energy constraint, since they can inject 
energy into the system independently of the electromagnetic excitation 
\cite{Serra2023}. Thus, such systems may have 
$\partial [\omega X(\omega)]/\partial \omega \le 0$, indicating that, during the transient 
process, the space-time surface emits more energy than it receives from the excitation. In this state, the system can provide gain in the electromagnetic response. Figure \ref{fig:reactance}(b, right panel) shows regions with negative stored energy.

\section{Electromagnetic Response of a Space-Time Grooved Surface}

Now that the space-time grooved surface has been homogenized, we proceed to analyze its electromagnetic features, including its scattering parameters, dispersion, and supported guided modes. This characterization allows us to link the effective constitutive parameters of the homogenized model with observable wave phenomena such as amplification and one-way propagation.

\subsection{Reflection Coefficient}

In the laboratory frame, the reflection coefficient 
$R$ can be calculated by means of the equivalent space-time surface impedance $Z_s$ that models the space-time grooved metal surface:
\begin{equation} \label{R}
    R = \frac{Z_s - Z_i}{Z_s + Z_i}\, ,
\end{equation}
where $Z_i$ is the impedance associated with the incident wave. In the TM case, it is given by
\begin{equation} \label{Zi_TM}
    Z_i = \frac{k_y}{\varepsilon_0 \omega}\,,
\end{equation}
where $k_y = \sqrt{(\omega/c)^2 - k_x^2}$ is the propagation  constant in the air region calculated in the laboratory frame. 

Substituting Eqs.~\eqref{Zs_TM} and \eqref{Zi_TM} into \eqref{R} and simplifying the resulting formula, we obtain an expression for the reflection coefficient:
\begin{equation} \label{R_TM}
    R = \frac{j\left( \omega - vk_x \right)  \tan \left( \frac{l}{c} [\omega - vk_x] \right) - k_y c}{j\left( \omega - vk_x \right)  \tan \left( \frac{l}{c} [\omega - vk_x] \right) + k_yc}.
\end{equation}

An interesting point to note is that, because $Z_s$ is purely imaginary, any propagating incident wave, for which $Z_i$ is real-valued, is reflected with unit amplitude. Thus, even though the space-time modulation is non-Hermitian and, as discussed in the previous subsection, gain can in principle be extracted from the space-time surface, this is not visible in the reflection response to propagating plane waves. The reason is that the non-Foster behavior occurs only in the spectral range $k_x > \omega/v$, which cannot be accessed by a propagating incident plane wave in the subluminal regime considered here, since $|k_x| < \omega/c < \omega/v$. It is therefore natural to investigate scenarios involving guided modes, which are not constrained by $|k_x| < \omega/c$ and hence may potentially unlock the gain. This is done in the following subsection.

\subsection{Guided Modes}

The dispersion equation that describes the guided modes supported by a single space-time surface can be found from the transverse resonance condition, \linebreak $Z_i + Z_s = 0$, applied at the spatial discontinuity $y=0$:
\begin{equation}
    \frac{k_y}{\varepsilon_0 \omega} + j Z_0 \left( \frac{\omega - vk_x}{\omega} \right)  \tan \left( \frac{(\omega - vk_x) l}{c} \right) = 0\, .
\end{equation}
It is instructive to consider first the quasi-static regime. For long wavelengths, we can use the approximations $k_y \approx -j |k_x|$ and $\tan (\alpha) \approx \alpha$. This reduces the dispersion equation to
\begin{equation}
    -|k_x|c + Z_0 \varepsilon_0 l (\omega - vk_x)^2 \approx 0\, ,
\end{equation}
which has the two analytical solutions
\begin{equation} \label{SolQSTM_1Bed}
    \omega(k_x) \approx vk_x \pm \frac{c}{\sqrt{l}} \sqrt{|k_x|}\, . 
\end{equation}

Strikingly, the dispersion remains real-valued for finite $v$, indicating that, despite the time modulation, the system does not exhibit instabilities. In fact, within the quasi-static description, the effect of the modulation is simply to imprint a Doppler shift on the dispersion of the unmodulated system. At first sight, this stability may seem puzzling, since the guided modes can potentially access the non-Foster regime. However, the reason  stability is preserved is simple: the effective surface impedance $Z'_s$ [Eq. \eqref{Zs_comoving}] in the comoving frame satisfies the usual passivity conditions in that frame. Hence, the dispersion calculated in the comoving frame is necessarily real-valued. Since the lab-frame dispersion is obtained from the comoving-frame dispersion through a Doppler transformation, it inherits this property.

The group velocity of the guided modes can be found as $v_{g,\pm} = d\omega/dk_x = v \pm c/(2\sqrt{l |k_x|})$. Thus, for a fixed velocity $v$, it is possible to find the minimum value of $k_x$ that ensures \emph{unidirectional propagation} ($v_{g,\pm}>0$) in the positive $+x$ axis: $k_x > c^2/(4lv^2) > 0$. This propagation asymmetry  is induced by the 
synthetic motion of the grooved metal surface, similar to the drift-induced unidirectionality reported for biased graphene sheets \cite{Morgado2018}. 

\begin{figure}[!t]
	\centering	\subfigure[\hspace{-1cm}]{\includegraphics[width=0.9\columnwidth]{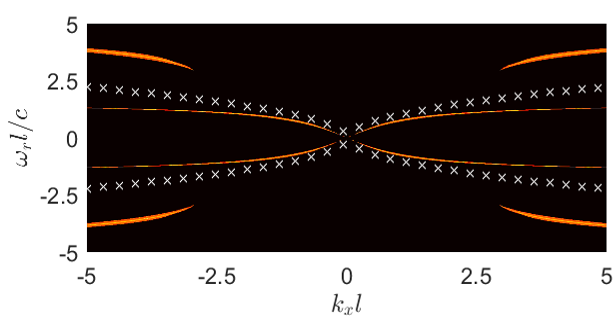}}
    \subfigure[\hspace{-1cm}]{\includegraphics[width=0.9\columnwidth]{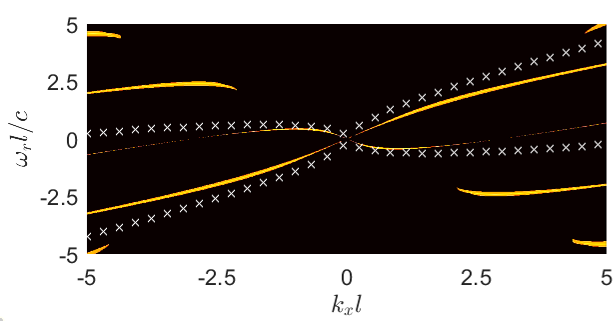}}
    \subfigure[\hspace{-1cm}]{\includegraphics[width=0.9\columnwidth]{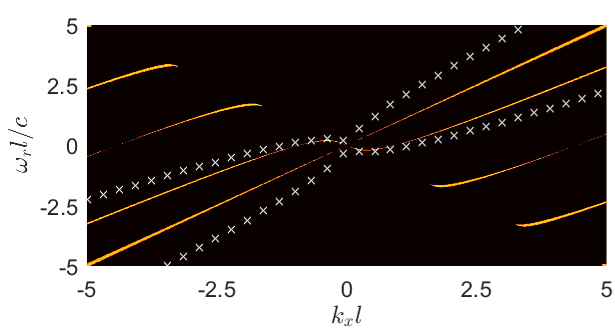}}
	\caption{Dispersion diagram showing the guided modes in a single space-time grooved surface for the velocities: (a) $v= 0$, (b) $v=0.4c$, (c) $v=0.9c$. The colored plot represents the numerical general solution and the white crosses represent the analytical quasi-static solution.} 
	\label{OneBed}
\end{figure}

Figure \ref{OneBed} represents the dispersion diagram of the guided modes for three different  velocities: \linebreak $v= \{0, 0.4, 0.9 \}c$. The solid lines represent the solution of the exact dispersion equation, whereas the white crosses represent the two analytical quasi-static solutions given in \eqref{SolQSTM_1Bed}. As seen, the quasi-static solutions match very well the full numerical response at the lowest frequencies and, additionally, for low velocities. 

Large values of $\omega$ and $v$ cause the argument of the tangent function in Eq. \eqref{Zs_TM} to become large, and the quasi-static solution fails in that regime. The periodic nature of the tangent function creates multiple replicas of the low-frequency branches of the guided modes, i.e., higher-order modes (see Figure \ref{OneBed}). The space-time modulation tends to make the higher-order modes unidirectional.

When the grooved metal surface is static [$v=0$, Figure \ref{OneBed}(a)], the dispersion diagram is perfectly symmetric, indicating that the structured platform is reciprocal. The synthetic motion induces a non-reciprocal response, which is manifested in the observed asymmetry in Figs. \ref{OneBed}(b)-(c). This implies, as already suggested by Eq. \eqref{SolQSTM_1Bed}, that it is possible to force the guided modes to propagate mostly along a single direction. For example, consider the example in Figure \ref{OneBed}(c). In this case, the quasi-static solution predicts that modes with wavenumbers such that $k_x l > 0.31$ propagate all along the $+x$ direction (for positive frequencies $\omega$). This analytical result is confirmed by the full numerical response represented in the colored plot of Figure \ref{OneBed}(c). 

The unidirectional behavior of the space-time grooved surface provides a pathway for engineering one-way propagation channels. The required space-time modulation can be implemented electrically, for example, by using synchronized switches that alternately ground or open the waveguide walls, thereby emulating the motion of the structure \cite{Moussa2023, Xiong2025}. Such implementations may also enable access to synthetic superluminal regimes, opening additional possibilities for wave control.

\section{Two-Facing Space-Time Grooved Metal Surfaces}


As discussed in the previous sections, the space-time-modulated grooved surface can provide a gain response, as confirmed by the fact that $\partial[\omega X(\omega)]/\partial \omega$ can be negative. Yet, the previous analysis shows that, despite this property, the dispersion of the guided modes remains real-valued. Next, inspired by the instabilities observed between moving bodies in shear motion \cite{Silveirinha2014}, we propose a strategy to extract gain from the system. Specifically, we consider a cavity formed by two periodically grooved surfaces subject to different modulation velocities. The motivation for considering this configuration is that it does not admit a global comoving frame. Consequently, there is no reference frame in which the material responses of both space-time gratings are simultaneously passive. As shown below, this property is the key to unlocking the gain.

\begin{figure}[!t]
	\centering	\subfigure{\includegraphics[width=0.9\columnwidth]{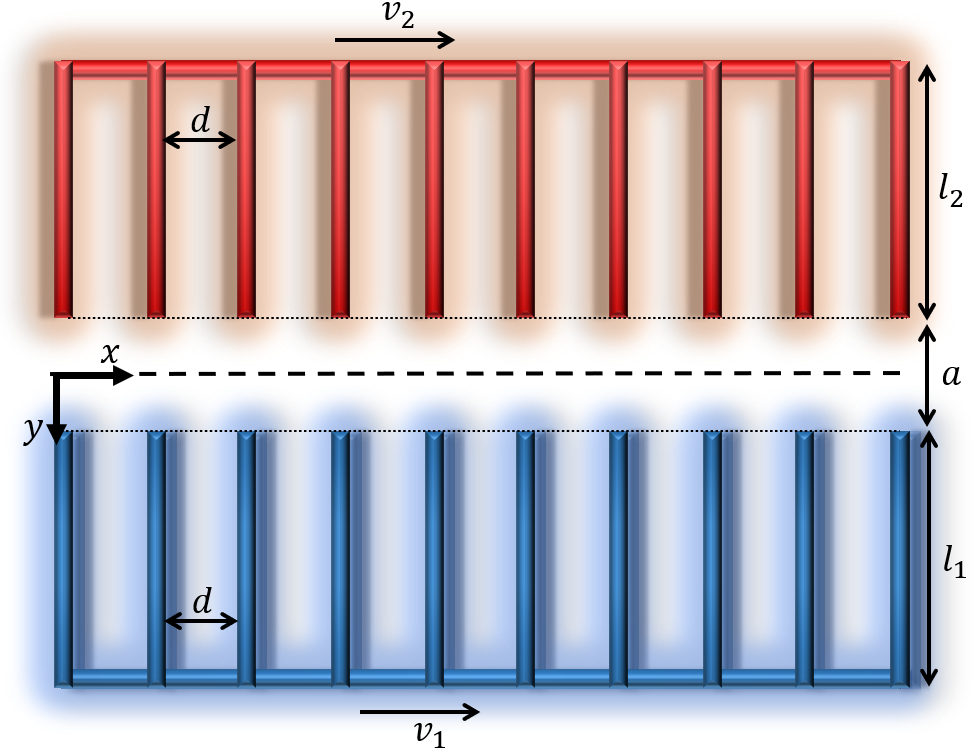}
	} 
	\caption{Sketch of two identical space-time grooved metal surfaces separated a distance $a$. The space-time structures have different dimensions, $l_1$ and $l_2$, and modulation speeds, $v_1$ and $v_2$.  } 
	\label{twobednails}
\end{figure}

The detailed geometry is shown in Figure \ref{twobednails}. The velocities of the two space-time-modulated surfaces are $v_1$ and $v_2$, and they are separated by a distance $a$.

Let us first consider the case where the two grooved surfaces are closely spaced ($k_x a \ll 1$). Then,  the application of the transverse resonance condition  [$Z_{s}(v=v_1, l=l_1) + Z_s(v=v_2, l=l_2) = 0$, $Z_s$ given by eq. \eqref{Zs_TM}]  yields the following dispersion equation for the guided modes:
\begin{multline}\label{dispcavity}
   (\omega - v_1k_x) \tan \left( \frac{(\omega - v_1k_x) l_1}{c} \right) \\ + (\omega - v_2k_x) \tan \left( \frac{(\omega - v_2k_x) l_2}{c} \right) = 0\, . 
\end{multline}

The previous dispersion equation lacks analytical solution in the general case, thus it must be solved numerically to find the values of $\omega(k_x)$. Nonetheless, some approximation can be made in the quasi-static regime, similar to what was done in  Section IV. The two analytical solutions found in the quasi-static regime are 
\begin{equation}  \label{SolTMAna_2Bed}
    \omega(k_x) \approx  \left(\frac{l_1 v_1 + l_2 v_2}{l_1 + l_2}\right) k_x \pm j \left( \frac{|v_1 - v_2| \sqrt{l_1 l_2}}{l_1 + l_2} \right) |k_x| \, .
\end{equation}

Importantly, quite different from Section IV, the quasi-static solutions are \emph{complex-valued}. In particular, for the time-variation $\mathrm{e}^{j \omega t}$ followed in the manuscript, the positive ($+$)  and negative ($-$) imaginary frequency solutions are associated with attenuation and amplification over time, respectively. Note that this effect is only possible when the two space-time surfaces are modulated with different velocities. 

\begin{figure}[!t]
	\centering	\subfigure[\hspace{-1cm}]{\includegraphics[width=0.9\columnwidth]{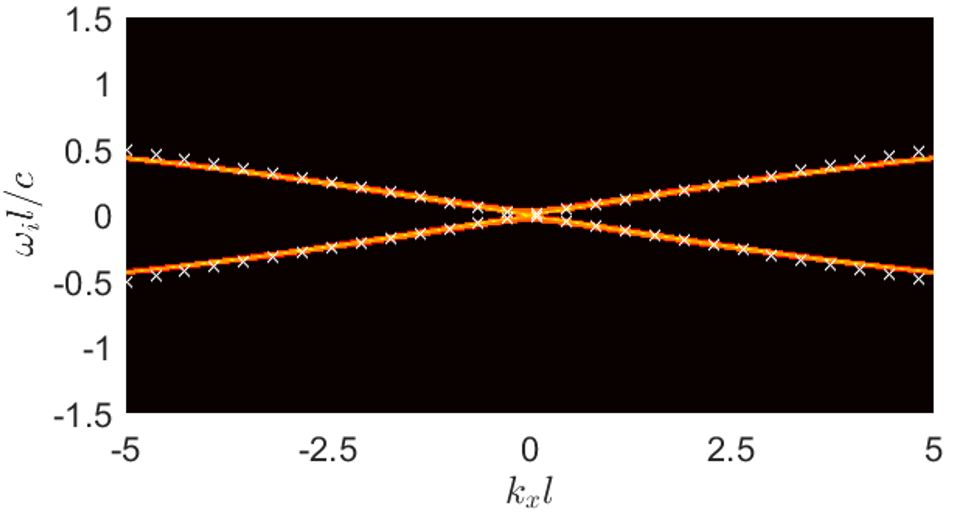}}
    \subfigure[\hspace{-1cm}]{\includegraphics[width=0.9\columnwidth]{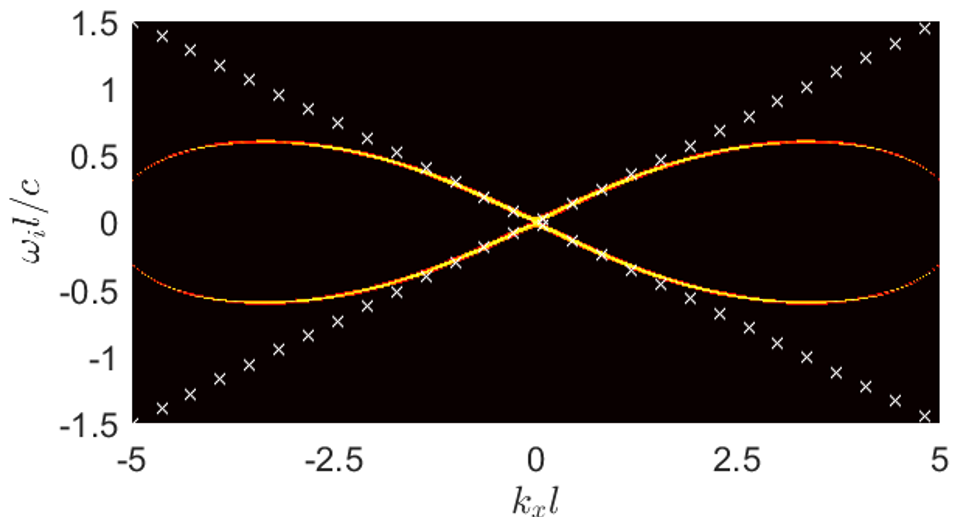}}
	\caption{Dispersion diagram showing the dissipation rate $w_i$ for the cavity modes in a scenario with two closely-spaced identical grooved surfaces that are modulated with the same synthetic velocity $v$ but in opposite directions. The considered velocities are: (a) $v= 0.1c$, (b) $v=0.3c$. The colored plot represents the numerical general solution and the white crosses represent the analytical quasi-static solution.} 
	\label{TwoBed}
\end{figure}

In order to simplify the discussion regarding amplification, let us consider a scenario with two identical grooved surfaces ($l_1 = l_2 = l$) that move with the same speed but in opposite directions ($v_1 = -v_2 = v$). Note that this case, the complex-valued analytical solutions for $\omega(k_x)$ in Eq.~\eqref{SolTMAna_2Bed} reduce to pure imaginary terms; namely, $\omega(k_x) = \pm j 2|v||k_x|$. This quasi-static expression suggests that the trend followed by the imaginary frequencies should be linear close to the origin. 

Figure \ref{TwoBed} shows the full solution for $\omega(k_x)=\omega_r+j \omega_i$, computed after numerically solving Eq. \eqref{dispcavity} for the case described above. Only imaginary frequencies $\omega_i$ are plotted since, as suggested by the quasi-static expression, no real frequencies are observed. The full numerical solution (colored line) is represented together with the quasi-static approximation (white crosses) given by Eq.~\eqref{SolTMAna_2Bed}. Good agreement is observed between both at the lowest frequencies. Solutions $\omega$ with a negative imaginary part are found in the dispersion diagram, indicating that the space-time-modulated system cavity supports an electromagnetic instability. As mentioned above, negative frequencies $\omega_i$ are associated with light amplification with the adopted time-harmonic convention. 

In the present example, with $v_1=-v_2$, the instability is inherently static ($\omega_r=0$). However, a simple analysis of Eq.~\eqref{dispcavity} shows that, in the general case, the dispersion satisfies
\begin{equation}  \label{dispgen}
    \omega(k_x) = \frac{v_1+v_2}{2} k_x +\omega^{\rm asym}(k_x),
\end{equation}
where $\omega^{\rm asym}(k_x)$ is the dispersion for the antisymmetric case ($v_1=-v_2$) with $v=(v_1-v_2)/2$ discussed above. This result shows that, by tuning $(v_1+v_2)/2$, one can shift the frequency $\omega$ at which the instability is strongest. Note that, from Figure~\ref{TwoBed}(b), the strongest instability occurs at a finite wave number $k_x$. In the example of Figure~\ref{TwoBed}(b), the strongest gain occurs for $k_x l \approx 3.35$.

In addition to this, regimes where the gap thickness $a$ becomes non-negligible ($k_x a \sim 1$) may also be analyzed using the same methodology developed in this section. Although the essential physical mechanisms associated with wave instabilities, amplification, and non-reciprocal or one-way propagation are already captured in the closely spaced limit ($k_x a \ll 1$), the gap size $a$ introduces an additional degree of freedom into the space-time-modulated system. In particular, $a$ strongly influences the near-field coupling between the two grooved surfaces. Similar to moving media \cite{Silveirinha2014}, the interaction responsible for the instability is mediated by evanescent surface modes whose amplitudes decay exponentially across the gap. Consequently, increasing $a$ reduces the overlap between the surface waves supported by each interface, thereby weakening the electromagnetic coupling and suppressing the instability growth rate. For sufficiently large gaps, the surface wave mediated interaction may become too weak to sustain amplification or spontaneous light generation.

\section{Conclusion}

In this paper, we studied space-time-modulated gratings formed by periodically grooved metal surfaces. It was underlined that although such systems are reminiscent of moving gratings, their electromagnetic response is different. Indeed, in a physically moving grating the PEC boundary condition is enforced in the comoving frame, whereas in a space-time-modulated system it is enforced directly in the laboratory frame.

We analyzed the homogenized response of these space-time structures by first studying a modulated parallel-plate waveguide using a Lorentz transformation. The dominant mode shows that one plate behaves as an energy source while the other acts as an energy sink. Then, by using transmission-line theory, we derived an equivalent surface reactance for the grooved surface. The obtained reactance exhibits non-Foster dispersion, including regions where $\partial(\omega X)/\partial \omega < 0$, indicating negative stored energy and the possibility of gain extraction.

Subsequently, the developed theory was applied to study effects such as light amplification and unidirectional propagation. Numerical results show that, for sufficiently large modulation velocities, the guided modes become predominantly unidirectional while their dispersion remains real-valued. Thus, the non-Hermitian nature of the space-time surface does not necessarily imply temporal amplification or attenuation.

Finally, we analyzed a cavity formed by two facing grooved surfaces modulated at different velocities. In this case, the cavity-mode dispersion can become complex, enabling electromagnetic instabilities and temporal amplification. Overall, these systems can support predominantly one-way propagation and enable electromagnetic amplification by extracting energy from the space-time modulation.

\begin{acknowledgments}
A. A.-A. and C. M. were supported in part by the Grant PID2024-155167OA-I00 funded by MICIU/AEI/10.13039/501100011033/FEDER, UE. A. A.-A. also acknowledges funding from the BBVA Foundation's Leonardo Grant for Scientific Research and Cultural Creation 2025. The BBVA Foundation is not responsible for the opinions, comments, and content included in the project and/or the results derived from it, which are the sole and absolute responsibility of their authors. C. M. was also supported in part by Consejería de Universidad, Investigación e Innovación of Junta de Andalucía through grant EMERGIA 23-00235.  M.G.S. was partially funded by the Simons Foundation (award SFI-MPS-EWP-00008530-10), and by national funds through FCT – Fundação para a Ciência e a Tecnologia, I.P., and, when eligible, co-funded by EU funds under
project/support UID/50008/2025 – Instituto de Telecomunicações, with DOI identifier https://doi.org/10.54499/UID/50008/2025.
\end{acknowledgments}

\appendix
\section*{Appendix. Lorentz Transformations}

Throughout the article, Lorentz transformations are employed to analyze the grooved surfaces in both the laboratory frame (unprimed notation) and the comoving frame (primed notation). For a synthetic uniform velocity $\mathbf{v} = v\hat{\mathbf{x}}$ directed along the $x$-axis, the corresponding transformations are summarized below \cite{KongBook}.

Spectral relations ($\gamma = 1 / \sqrt{1 - (v/c)^2}$):
\begin{equation}
\omega' = \gamma (\omega - v k_x), \qquad k_x' = \gamma \left(k_x - \frac{v}{c^2} \omega\right).
\end{equation}
\begin{equation}
    k_y' = k_y, \qquad k_z' = k_z.
\end{equation}

Field relations:
\begin{equation} \label{fields1}
E_x' = E_x, \qquad B_x' = B_x,
\end{equation}
\begin{equation} \label{fields2}
E_y' = \gamma \left(E_y - vB_z\right), \qquad B_y' = \gamma \left(B_y + \frac{v}{c^2}E_z\right),
\end{equation}
\begin{equation} \label{fields3}
E_z' = \gamma \left(E_z + vB_y\right), \qquad B_z' = \gamma \left(B_z - \frac{v}{c^2}E_y\right).
\end{equation}
The inverse field relations are obtained by simply replacing $v\rightarrow -v$:
\begin{equation} \label{fields4}
E_y = \gamma \left(E_y' + vB_z'\right), \qquad B_y = \gamma \left(B_y' - \frac{v}{c^2}E_z'\right),
\end{equation}
\begin{equation} \label{fields5}
E_z = \gamma \left(E_z' - vB_y'\right), \qquad B_z = \gamma \left(B_z' + \frac{v}{c^2}E_y'\right).
\end{equation}


\bibliography{apssamp}%

\end{document}